\documentclass[twocolumn,pre,showpacs,preprintnumbers,amsmath,amssymb,nofootinbib,superscriptaddress,floatfix]{revtex4-1}
\usepackage{amsfonts}
\usepackage{amsmath}
\usepackage{amssymb}
\usepackage{graphicx}
\usepackage{dcolumn}
\usepackage{bm}
\usepackage{comment}
\usepackage{longtable}
\usepackage[usenames, dvipsnames, table]{xcolor}
\newcommand{\lp}{\left (}
\newcommand{\rp}{\right )}

\newcommand{\ra}{\rightarrow}
\newcommand{\R}{{\mathbb{R}}}
\newcommand{\ev}[1]{\left \langle #1 \right \rangle}

\newcommand{\beq}{\begin{equation}}
\newcommand{\eeq}{\end{equation}}

\newcommand\mnras{MNRAS}%
\begin{document}
%\preprint{}
%\title{A Unifying Theory for Scaling Laws of Human Populations}
\title{Zipf's Law from Scale-free Geometry}
%\author{}
%\thanks{}
%\affiliation{}
\author{Henry W. Lin}
%\email[]{}
\affiliation{Harvard College, Cambridge, MA 02138, USA}
\author{Abraham Loeb}
\affiliation{Institute for Theory \& Computation, Harvard-Smithsonian Center for Astrophysics, 60 Garden Street, Cambridge, MA 02138, USA}
\date{\today}
\begin{abstract}
The spatial distribution of people exhibits clustering across a wide range of scales, from household ($\sim 10^{-2}\, \text{km}$) to continental ($\sim 10^4\, \text{km}$) scales. Empirical data indicates simple power-law scalings for the size distribution of cities (known as Zipf's law) and the population density fluctuations as a function of scale. %We construct a simple analytical model that explains all three of these scaling laws based on a single unifying principle involving the random spatial growth of clusters of people on all scales. 
Using techniques from random field theory and statistical physics, we show that these power laws are fundamentally a consequence of the scale-free spatial clustering of human populations and the fact that humans inhabit a two-dimensional surface. In this sense, the symmetries of scale invariance in two spatial dimensions are intimately connected to urban sociology. We test our theory by empirically measuring the power spectrum of population density fluctuations and show that the logarithmic slope $\alpha = 2.04 \pm 0.09$, in excellent agreement with our theoretical prediction $\alpha = 2$. The model enables the analytic computation of many new predictions by importing the mathematical formalism of random fields. %instead of traditional discrete graph techniques.
\end{abstract}
\pacs{}%89.65.Gh; 89.75.Hc; 05.45.Tp; 02.50.Sk}
%\keywords{}
\maketitle
\section{Introduction}

Human populations exhibit remarkably simple properties given the complexity of socioeconomic interactions between humans and their environments\cite{betten13}. One such example is the well known {\it Zipf's law}\cite{zipf} for cities: the rank of a city is inversely proportional to the number of people who live in the city. If the most populous city in the United States has a population of $N_\text{max,US} \sim 8 \times 10^6$, the second most populous city will have a population of ${1 \over 2}N_\text{max,US} \sim 4 \times 10^6$, the third ${1 \over 3}N_\text{max,US} \sim 2.7 \times 10^6$, and so forth. This simple relation fits empirical data extremely well\cite{gab99, jiang11}. A mathematically equivalent formulation of Zipf's law is that the underlying distribution of cities follows a power law\cite{newman}; namely, the probability that a city has a population $N$ scales as $1/N^2$. 

%Another remarkable scaling law\cite{lib05,kumar05} states that the probability a given person $A$ befriends another person $B$ is inversely proportional to the number of people $N_{B,A}$ geographically closer to $A$ than $B$. $N_{B,A}$ is commonly referred to as the rank of $B$, so we will refer to this relationship as the {\it inverse-rank friendship law}. In the limiting case where people are distributed uniformly over a 2-dimensional surface, the probability will scale as $\sim 1/r^2$ where $r$ is the separation between people. The general relation, which leads to a complicated spatial dependence when clustering is present, appears to be valid at least on the scale of towns and cities\cite{lib05}.

%On the theoretical side, recent work\cite{pan13} used the inverse-rank friendship law to explain a variety of urban characteristics such as the length of phone calls made, the spread of sexually transmitted diseases, and economic activity, but did not comment on the theoretical origin of the law. On the other hand, 
%
The remarkable simplicity and empirical success of Zipf's law have attracted significant theoretical attention and debate\cite{mars98, gab99, baek11}, though there is no consensus on the origin of Zipf's law. Existing work treats cities as the fundamental entities of the theory, with population as a property of each city. For example, Gibrat's law applied to cities\cite{gibrat,samuels,gab99}, which states that the fractional growth rate of a city is independent of its population, will drive the distribution of city populations to a log-normal distribution. The tail of the log-normal distribution then gives rise Zipf's law.

Our approach is conceptually different: we treat the population density as the fundamental quantity, thinking of cities as objects that form when the population density exceeds a critical threshold. The situation is therefore conceptually and mathematically analogous to the formation of galaxies in the universe, where non-linear gravitational collapse occurs when the matter density exceed some critical value. Our conceptual advance here is also a practical one, since we can apply the mathematical tools developed for analyzing random fields\cite{ps74} to the problem at hand.

Before proceeding with a technical derivation of our results, let us briefly summarize them. The starting point is to model human population density as a random function of spatial position. A function of spatial position is a {\it field}, and thus human population density will be modeled as a {\it random field} (for a review of relevant topics in random fields, see \cite{randomfields,kardar} and especially \cite{bardeen86}). To lowest order, a single random variable in elementary statistics is characterized by a mean and a variance. A random field may be regarded as a higher dimensional generalization of a single random variable. By analogy, a random field is characterized by a mean and a {\it power spectrum}, which can be thought of as a generalization of variance. The power spectrum gives the amount of fluctuations of the field as a function of scale. To derive the form of the power spectrum for human population density, we invoke scale-invariant random growth, similar in spirit to Gibrat's law. 

We move on to derive Zipf's law. Our derivations involve the simple assumption that some cities emerge above some critical population density threshold. %For the derivation of Zipf's law, cities emerge, and in the friendship law, communities of friends emerge. 
To count the number of cities in our model, one must answer the following mathematical question: given a random field characterized by a power spectrum, how often does the random field take on values greater than a certain threshold? This is a frequently asked question in the context of cosmology, and the Press-Schechter (PS) formalism allows us to analytically compute the answer.

%sing both simple dimensional analysis and effective field theory approach
We demonstrate that our derivation of Zipf's law is more general than the motivating random growth model; we argue that the only key ingredient is scale invariance in two spatial dimensions. 
In other words, whereas previous work tends to focus on how Zipf's law emerges from concrete models, we argue that Zipf's law naturally occurs in a very large class of statistical models. In the language of statistical physics\cite{kardar}, the existence of Zipf's law is only a function of the universality class of the statistical model; it is independent of the ``microscopic'' details of the system's dynamics which are undoubtedly complex in the case of human populations. 

\section{Derivation of Zipf's law}
We now proceed with the detailed derivation. To start, consider the human population density $\rho$ as a function on $\R ^2$, the 2D Euclidean plane. Since we will be interested in regions much smaller in size than the radius of the Earth, we will ignore the effects of curvature. The fluctuations relative to the average population density $\delta({\bf x}) \equiv [(\rho({\bf x})/\bar{\rho}) - 1]$ can be expanded in Fourier modes 
$$\delta({\bf x}) = \frac{1}{2\pi}\int d^2 k \, \delta_{\bf k} \, e^{-i{\bf k}{\bf x}}.$$ Up to a conventional normalization factor of $2\pi$, this equation simply rewrites the population fluctuations as a sum of plane waves $e^{-i {\bf k x}}$, each weighted by a factor $\delta_{\bf k}$. Since the left hand side is a random variable, the right hand side must also be a random variable; since every term except for $\delta_{\bf k}$ on the right hand side is manifestly deterministic, $\delta_{\bf k}$ must be a continuum of random variables, with one random variable for each wave vector $\bf{k}$. Just as an ordinary random variable is characterized by a variance, each $\delta_{\bf k}$ is characterized by a number $P({\bf k})$ called the power spectrum, which is defined as
\beq
\ev{\delta_{\bf k} \, \delta_{\bf k'}^*} = (2\pi)^2 \delta_D^2 ({\bf k}-{\bf k'}) P(k),
\eeq where $\delta_D$ is the Dirac delta function (not to be confused with the fractional over-density $\delta({\bf x}$). By assuming rotational symmetry, the power spectrum becomes a function only of magnitude $P({\bf k}) = P(k)$. Equation (1) makes precise the statement that the power spectrum $P(k)$ quantifies the amount of statistical fluctuations associated with a given frequency $k$.

%The power spectrum quantifies the variance %the factor of $1/2\pi$ is a conventional normalization factor,
%The amplitude of the population density fluctuations with wavenumber $k$ (or an associated length scale $\sim 1/k$) is given by the cylindrically averaged power spectrum $P(k)$, defined by 

It is conventional to define a dimensionless power spectrum in the number density $\Delta^2(k) \equiv k^2 P(k)/(2\pi)$, which represents the typical (squared) fractional over-density of people $\lp \delta \rho/\rho\rp^2$ on the spatial scale $\sim 1/k$. To make further progress, we must fix the functional form of $\Delta(k)$ by some theoretical principle. To this end, consider an over-density of size $\sim 1/k$. At a discrete time step, this over-density might grow or shrink in spatial coverage. As a concrete example, consider a collection of farms (with a characteristic population density of a few people per typical farm area) in otherwise relatively uninhabited countryside. At each time step, a farm could be added or destroyed. In this way, our unifying principle of random walkers is conceptually similar to previous work on the random growth of firms \cite{zamb15}. Therefore, the spatial size of the over-density might grow or shrink, while $\delta \rho/\rho$ (a number associated with farms) will be held constant.
More precisely, we define a monotonically decreasing function $X(k)$ such that $\lim_{k \ra \infty} X = 0$, which quantifies the spatial extent of an over-density. This function might represent the area of the over-density $X(k) \propto 1/k^2 $ or its perimeter $X(k) \propto 1/k$, but our derivation will not depend on the detailed form of $X$. We can then perform a change of variables and view $\Delta(k)$ as a function of $X$: $\Delta(X(k)) = \Delta(k)$. The unifying principle is that all over-densities can grow or shrink spatially, executing a random walk in $X$. This process can continue until the overdensity disappears $(X = 0)$, or the over-density takes up some maximum $X_\text{max}$, where $X_\text{max} \equiv X(k_\text{min})$ is set by the continental length scale $\sim 1/k_\text{min}$. For a large ensemble of over-densities, this is a diffusion-like process with reflecting boundary conditions obeying
%
 %and define a monotonically decreasing (but otherwise arbitrary) function $X(k)$.  Our key principle is as follows. Therefore, the value of $k$ that corresponds to the given over-density will change. In particular, we assume that $X(k)$ corresponding to the given over-density increases or decreases by one unit with equal probability. We can state this in more concrete terms. Imagine a group of farms out in the countryside, with each farm contributing a characteristic population over-density relative to the uninhabited countryside.  At a given time step, new farms might pop up next to existing farms increasing $X$ or farms might get replaced by either empty space or more densely populated urbanized areas decreasing $X$. Thus the ``clumpiness" $\Delta(k)$ on the length scale $
%\sim 1/k$ executes a random walk. This process can continue until there are no farms, or the collective farmland takes up some maximum $X_\text{max}$, where $X_\text{max} \equiv X(k_\text{min})$ is on the order of the surface area of the continent if $X \sim 1/k^2$. More precisely, the over-density takes a random walk in $X$ with reflecting boundaries at 0 and $X_\text{max}$. For a large ensemble of over-densities, this is a diffusion-like process obeying
\begin{equation}
\frac{\partial \Delta}{\partial t} = D \frac{\partial^2 \Delta}{\partial X^2} 
\end{equation}
with some diffusion constant $D$. We are only interested in the late-time behavior of equation (2). Any initial conditions will relax to the steady-state solution $\Delta(X) \ra$ constant for $0 \le X \le X_\text{max}$ on a timescale $T_\text{relax} \sim X_\text{max}^2/D$. We intuitively expect $T_\text{relax}$ to be reasonably short, since the geographic mobility timescale of $\sim 5$ yrs (in the United States, $\sim 35\%$ of people change residences within 5 years\cite{ihrke}) is considerably shorter than, say, the population growth timescale $\sim 30$ yrs set by the typical age of parenthood. Any initial conditions set by antiquity or perturbations to the system (e.g. catastrophic events that displace many people) should be quickly erased. We therefore predict that on sufficiently long timescales,
\begin{equation}
P(k) \propto k^{-2}.
\end{equation}
We test this prediction in Figure 1 against publicly available data from the Center for International Earth Science Information Network (CIESIN) and Centro Internacional de Agricultura Tropical (CIAT)\cite{url}. We find the best fit slope $P(k) \propto k^{-\alpha}$ to be $\alpha = 2.04 \pm 0.09$, where we have reported the $\pm 1 \sigma$ uncertainties. Our theoretical prediction is therefore in excellent agreement with observations across a broad range of spatial scales, from a few km to $\sim 10^3$ km.

\begin{figure}
\centering
\includegraphics[width = 0.45 \textwidth]{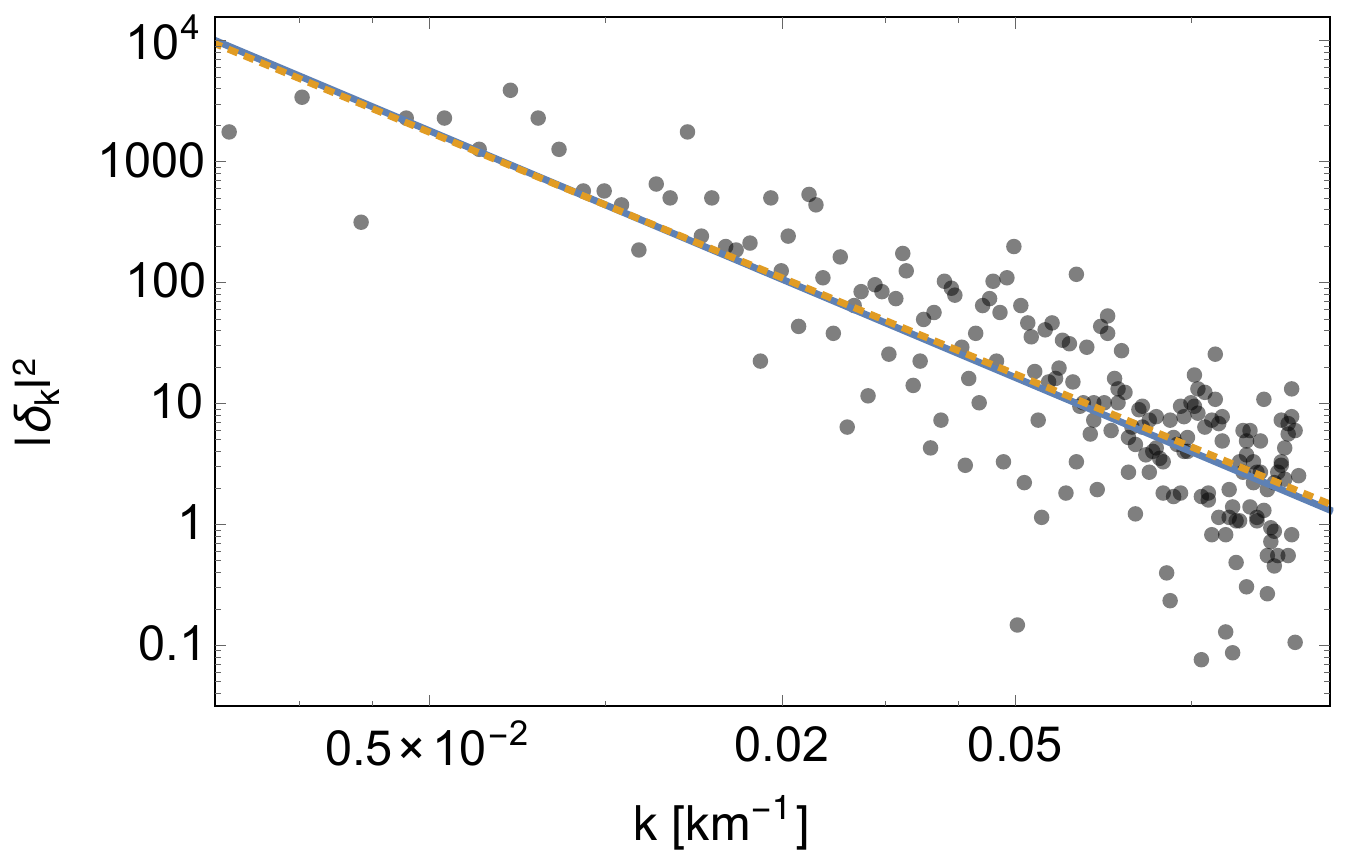}
\caption{(Color online). Empirically measured power spectrum $P(k) \propto |\delta_k|^2 \sim k^{-\alpha}$ of population density fluctuations as a function of the spatial wavenumber $k$. The best fit slope $\alpha = -2.04 \pm 0.09$ (solid blue line) is virtually indistinguishable from the predicted slope $\alpha = -2$ (dashed orange line). The data was obtained by taking the diagonal entries (to avoid anisotropy from rectangular gridding) of a discrete Fourier transform of a $1000 \times 1000$ arcmin$^2$ map of the population density of a section of the continental United States. The area was selected to minimize artifacts due to boundary conditions defined by lakes and oceans.}
\end{figure}

Before further developing the theory, a more intuitive derivation of $P(k)\sim k^{-2}$ is worth mentioning. Over a large range of length scales our model is scale free, implying $P(k)\sim k^{-\alpha}$ for some $\alpha$. In $d$ spatial dimensions, the left hand side of equation (1) has units of $k^{-2d}$, the Dirac delta function has units of $k^{-d}$, so $P(k)$ should have units of $k^{-d}$. Since there are no other dimensional parameters relevant to our theory (the diffusion constant has units of $[X]^2 T^{-1}$, but there are no other constant with units of time $T$), we must have $\alpha=d =2$ in two spatial dimensions. In this sense, geometry and scale invariance uniquely determines the slope of the power spectrum.

In fact, this simple argument demonstrates that $P(k)\propto k^{-2}$ is a universal feature of 2D models that have no parameters with units of length to some power. Effective field theory, a powerful technique for studying any statistical physics system, can be used to further sharpen this statement; this is done in Appendix D. The derivation in Appendix D provides perhaps the most rigorous way of justifying the statement that $P(k) \propto k^{-2}$ is a generic property of models with scale invariance in two dimensions, since effective field theory should capture the equilibrium properties of any statistical model on scales smaller than the system scale but sufficiently large such that densities can be approximated by smooth functions.%For example, a simple noisy spatial diffusion model which will also yield $P(k)\propto k^{-2}$ is presented in Appendix D.

{ With a power spectrum $P(k)$ in hand, it is possible to calculate the number of cities as a function of their population $N$. We picture cities of area $A$ as discrete objects which form when the population density as a function of spatial coordinates $\rho({\bf x})$, or equivalently $\delta({\bf x})$, averaged over an area $A$ surpasses a critical threshold, $\delta_C$. In other words, we choose the surface area $A$ such that the total integrated population $N = \int_{x \in A} \rho({\bf x}) \, d^2 x = \rho_C \times A$, where the critical density $\rho_C = \bar{\rho}(1 + \delta_c)$.

This is shown pictorially in Figure 2. [This assumption can be relaxed, allowing for the average population density of a city to vary systematically with size. In our model, this corresponds to a critical threshold that varies with $A$. In this case, the excursion set formalism can be used with a moving barrier\cite{sheth99}. We will ignore this subtlety, since Zipf's law is still obtained in the limit that $\delta_C \ll \sigma$. Also, since the numerical value of the threshold is not fixed, we could also consider the case where the threshold varies by country. Again, Zipf's law would be obtained for each country.] 

The counting of cities is now a well-posed question. Computationally, one could find the number distribution of cities with the following algorithm. Generate via a Monte Carlo procedure many realizations of the random field with mean 0 and power spectrum $P(k) = P_0 k^{-2}$. Find the regions where the random field exceeds a certain threshold. Measure the size of each region, and multiply the area of each region by the population density threshold; define this to be the population of each city.  Repeat for many Monte Carlo iterations, and then make a histogram of the size distributions of each region. One can verify numerically that the resulting number distribution $n(N)$ would scale approximately like
\beq
n(N) \propto N^{-2},
\eeq
where $N$ is the population of the city and $n(N)$ is the number density of cities of size $N$. However, we can in fact show {\it analytically} that the number distribution takes this form using the Press-Schechter (PS) formalism\cite{ps74}, traditionally used in the context of cosmology to predict the abundance of gravitationally-bound objects given a power spectrum of the fluctuations in the cosmic matter density. However, we emphasize that the formalism is in essence a purely statistical one, which does not require or employ any facts from cosmology. The excursion set formalism\cite{bond91} provides a more rigorous derivation, but the PS formalism has the benefit of simplicity. The end result is identical in either case. We provide a self-contained proof of equation (4) in Appendix A.}

By integrating equation (4) with respect to $N$, we find that the number of cities above a certain population threshold scales inversely with the population threshold. This statement is equivalent to Zipf's law: the rank of a city is inversely proportional to its size.
%%%%%%%%%%%%%%%%%%%%%%%%%%%%%%%%

%%%%%%%%%%%%%%%%%%%%%%%%%%%%%%%%%%%%%%%%
\begin{figure*}
\centerline{\includegraphics[width = .9 \textwidth]{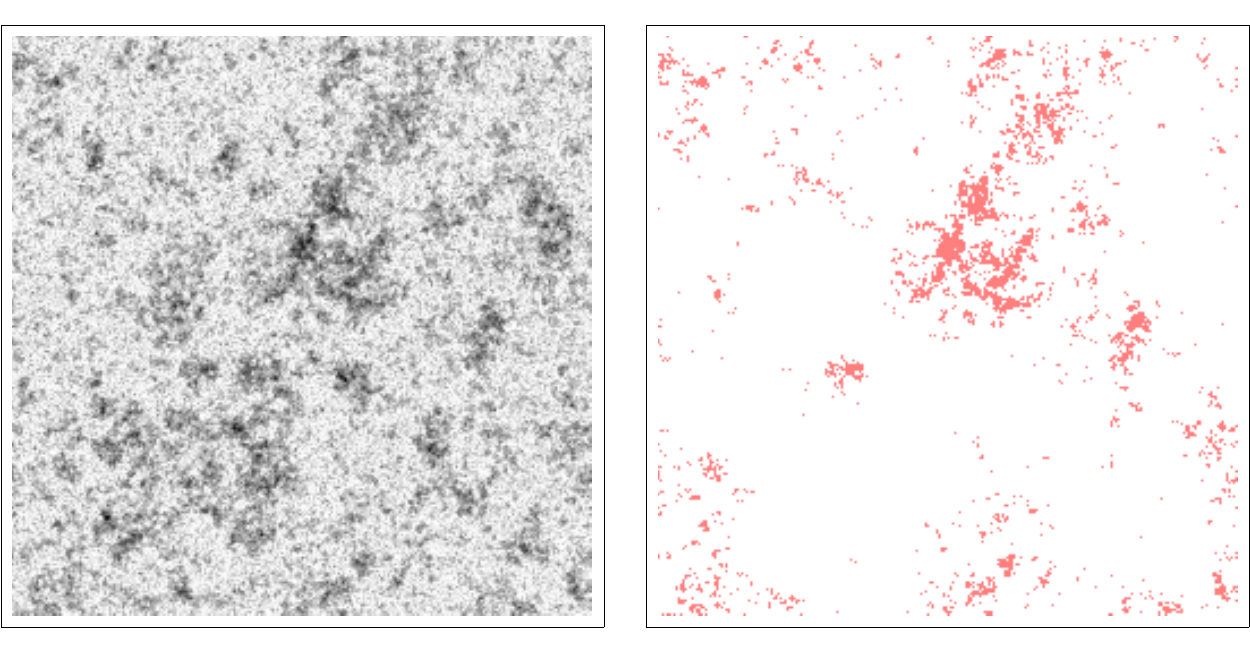}}
\caption{(Color online). Schematic illustration of our approach. On the left, a simulated population density map with a power spectrum $P(k) \propto k^{-2}$ is displayed. Darker pixels indicate higher population densities. On the right, we select and color pink all pixels above a certain population density threshold from the simulated map on the left. In our formalism, a city is identified with each pink cluster, appropriately smoothed on the length scale of the cluster. Scale invariance implies that cities of all sizes appear on the map, as confirmed by visual inspection. In our formalism, the statistical size distribution of pink cluster gives us the population distribution of cities, which agrees with Zipf's law.} %We use a similar approach for calculating the sizes of ``friendship" communities and the size distribution of epidemics.} %Plotted vertically is the smoothed spatial population density over a strip of the continental United States. A region of length $\Delta L$ where the average population density is greater than some threshold value is declared a city of size $\Delta L$. Given the power spectrum of population density fluctuations $P(k)$, we analytically compute the number distribution of cities. We use a similar approach for calculating the sizes of ``friendship" communities and the size distribution of epidemics.}
\end{figure*}
\section{Conclusion}
In summary, we have presented a derivation of Zipf's law and successfully predicted the power spectrum $P(k)$ of population density fluctuations in the continental US. These derivations stemmed from two fundamental ingredients: scale-invariance and 2D geometry. Remarkably, there is a wide range of possible models and an even wider range of initial conditions to which our results are insensitive. One such model involves random walks of the sizes of clusters of people on all scales, which can be viewed as a vast generalization of Gibrat's law. However, as we have emphasized, even this generalization is still a relatively specific example in the class of all models which will lead to Zipf's law.%However, there are many other possibilities, as emphasized in the dimensional the universality argument  Appendix D that would also.
This shows that the origin of these laws is fundamentally the scale-free nature of clustering in human populations. This is an appealing feature, enabling us to forgo any fine-tuning arguments in explaining the empirical data. %Future work here could include the calculation of a host of new observables such as the bias factor\cite{mo96} for the spread of epidemics or other social phenomena and refinements to the detailed theory of clustering based on additional data.

\begin{acknowledgments}
The authors would like to thank the anonymous referees for useful discussions. This work was supported in part by NSF grant AST-1312034.
\end{acknowledgments}

\begin{appendix}
%%%%%%%%%%%%%%%%%%%%%%%%%%%%%%%%%%%%%%%%%%%%%%%%%%%%%%%%%%%%%%%%%%%%%%%%
\section{The Press-Schechter formalism}
%%%%%%%%%%%%%%%%%%%%%%%%%%%%%%%%%%%%%%%%%%%%%%%%%%%%%%%%%%%%%%%%%%%%%%%%
{The Press-Schechter formalism (for a pedagogical overview of the PS formalism and its generalizations, see section 3.4 of \cite{loeb13}.) allows us to answer the well-posed question: given a random field with an associated power spectrum $P(k)$, how often does it exceed the threshold? More specifically, suppose there is a class of objects (e.g. cities) that form when the population density exceeds a certain threshold $\rho(x) > \rho_\text{threshold}$. Furthermore, let the size $R$ of each object be defined as the maximum radius $R$ such that the average population density $\rho_\text{circle,R}$ within a circle of radius $R$ centered on the object is given by 
\beq\rho_\text{circle,R} = \rho_\text{threshold}.\eeq
It is conventional to define a smoothed density field 
\beq
\delta_A({\bf x}) = \int d^2 k \, W_A(k) \,\delta_{\bf k} \, e^{-i{\bf kx}}/(2 \pi)^2,
\eeq 
where the low-pass window function $W_A(k) = 1$ if $k\le 1/\sqrt{A}$ and $W_A(k) = 0$ otherwise. This smoothed field is simply the original field $\delta(x)$ with the high-frequency fluctuations subtracted out, leaving behind the slowly-varying components.

For a fixed ${\bf x}$, $\delta_A({\bf x})$ is a random variable with probability distribution $p_A(\delta_A)$. The key insight of the PS formalism is to identify the fraction $f_A$ of people living in cities of area $A$ or larger with the cumulative probability $f = 2 \int_{\delta_C}^\infty p_A \, d \delta$. This is illustrated in Figure 2.
To make further progress, we must assume something about the functional form of $p_A$. The conventional PS formalism assumes that $p_A$ is a Gaussian with mean 0 and variance $\sigma^2(A) \equiv \int_{k_\text{min}}^{1/\sqrt{A}} d k \, k P(k)/\lp 2\pi\rp \propto \ln k/k_\text{min}$. If each Fourier mode is statistically independent of every other Fourier mode, the density field will be the sum of many independent Fourier modes and will therefore be approximately Gaussian. However, for the sake of generality, we will not assume that $\delta$ is normally distributed. Instead, we only assume that $p_A$ has a universal shape for all $A$. Since the mean of $\delta_A$ is zero for all $A$ by definition, and since for any random variable $\theta$ the associated standard deviation obeys $\sigma_{a\theta} = a \sigma_\theta$, this allows us to write $p_A(\delta) = g(\delta/\sigma(A))/ \sigma(A)$ for some general probability density function $g$. Differentiating $f_A$ yields $n(N)$, the number of cities on Earth's surface with population $N = \bar{\rho} A$ per unit area per unit population: 
\begin{equation}
n(N) =  -\nu g(\nu) \frac{\rho}{N}  \frac{d \ln \sigma}{d N} \propto \frac{1}{N^2} \frac{g(\nu)}{\ln(N_\text{max}/N)}, 
\end{equation}
where we have defined $\nu \equiv \delta_C/\sigma(N)$, the number of standard deviations associated with city formation. 
Note that for $\nu \ll 1$, $g(\nu)$ is a slowly varying function of $N$ for two reasons: the first derivative of $g$ around $\nu = 0$ is small for small deviations from the mode, and $\nu$ is only a weak function of $N$.  
Thus, equation (3) implies that the logarithmic slope $d \log n/d \log N$ tends to $-2$ in the limit of $N \ll N_\text{max}$. This limit is empirically justified, since even the largest cities in the world contain only $\sim 10^{-3}$ of the world's population. Hence we arrive at equation (4).

Although our results are largely independent of the exact form of $p_A(\delta)$, let us briefly comment on its possible form. If $p_A(\delta)$ deviates from a Gaussian, this implies that different Fourier modes in human population density are correlated, a generic result of non-linear interactions. Note, however, that $\delta \ge -1$ is strictly bounded from below, since human population density is always positive-definite: $\rho \ge 0$. Hence, $p_A$ cannot be exactly Gaussian. At some level, non-linear interactions must come into play. If the population density fluctuations were typically small $\delta \lesssim 1$, one might expect that a Gaussian distribution could be a good approximation; however, everyday experience tells us that population density fluctuations can be quite large. Indeed for New York City, $\delta \sim 300$. Hence, a theory of human population density growth must necessarily be a non-linear.}
\\

\section{Derivation of the inverse-rank friendship law}
As a second application of our formalism, let us derive the average number of friends a person has in a given region. We again adopt a simple model, where we define a region to be a {\it community} if the population density exceeds some critical value $\delta \ge \delta_c$. This defines geographic equivalence classes on the inhabited regions, such that every person is a member of a community. Since real-world social networks are highly clustered and only a small fraction of people serve as connections between communities of friends \cite{watts98}, this assumption should be a good approximation for our purposes, since the more complicated topology of real-world friendship networks will mainly affect higher order quantities that involve friends-of-friends and friends-of-friends-of-friends. Furthermore, we assume that the average number of friends $D$ a given person has is asymptotically independent of the size of the community. This second assumption is essentially the assertion of the existence of the famous Dunbar's number \cite{dunbar, gon11}, an upper limit on the number of people a given person can sustain social relationships with.

To compute the probability in the model, we consider two people $A$ and $B$ with $N_{AB}$ people closer to $A$ than $B$. If $A$ is a member of a community with size $N_c \gg N_{AB}$, $A$ and $B$ are almost certainly friends. On the other hand, if $A$ is in a community of size $N_c \ll N_{AB}$, it will be nearly impossible for $A$ and $B$ to be friends. There is thus a turnover scale at $\sim N_{AB}$ which dictates whether or not $A$ and $B$ will be friends; the probability is therefore determined by two independent events: the probability that $A$ is in a community of size $N$ greater than the turnover scale and the probability $p_f = D/N$ that $A$ and $B$ are friends given that $A$ and $B$ are in a community of size $N$, for large $N \gg D$. Since we know from the previous discussions that in such a model, the number density of communities scales asymptotically with $\propto 1/N^2$, and each community has $N$ people, the probability $p_c$ that a randomly chosen individual is in a community of size $N$ scales $\sim 1/N$. Hence,
\beq 
\begin{split}
p(N_{AB}) = \int g(N,N_{AB}) p_c(N) p_f(N) d N \\ \propto  \int_{N>N_{AB}} \frac{1}{N} \frac{D}{N} d N \propto \frac{1}{N_{AB}}
\end{split}
\eeq
%Let us now consider an arbitrary person in a region with a total of $N_R$ people. A randomly chosen person will have on average $N_f$ friends given by
%\begin{equation}
%N_{f} = \frac{1}{Z}\int_1^{N_R} N \, p_c(N) \, dN \propto \log N_R + {\cal O}\lp \frac{\log N_R}{N_R}\rp
%\end{equation}
%where $Z = 1-\int_{N_f}^\infty p_c(N) dN$. 
where $g$ has the properties that $0\le g < 1$, $g \approx 1$ for $N\gg N_{AB}$, and $g \ll 0$ for $N\ll N_{AB}$. The details of the function will depend on the geometry of the communities but do not concern us here as we are only interested in the scaling. We have thus derived the inverse-rank friendship law, previously proposed\cite{lib05} to fit empirical data. We stress that our derivation is based entirely on theoretical considerations and therefore provides an explanation for the ``physical'' origin of the law.
\\
%%%%%%%%%%%%%%%%%%%%%%%%%%%%%%%%%%%%%%%%%%%%%%%%%%%%%%%%%%%%%%%%%%%%%%%%
\section{Two-point correlation function}
%%%%%%%%%%%%%%%%%%%%%%%%%%%%%%%%%%%%%%%%%%%%%%%%%%%%%%%%%%%%%%%%%%%%%%%%
In this appendix, we analytically compute the two-point correlation function\cite{smith2014} $\xi(x-y) = \ev{\delta(x)\delta(y)}$, which is the inverse Fourier transform of the power spectrum. The correlation function will play an important role in Appendix D. Physically, the correlation function measures the degree to which the existence of an over-density or under-density at some position $\bf{x}$ increases the likelihood that an over-density or under-density will be found at $\bf{y}$. Assuming circular symmetry, the inverse Fourier transform is a Hankel transform of order 0:
\beq
\xi(r)=\int_0^\infty \frac{k \, d k}{2\pi} \, P(k) J_0(k r),
\eeq
where $J_0$ is the first Bessel function. Taking $P(k) = P_0 k^{-2}$ and a long-wavelength cutoff $k_m$ gives us an integral that can be written in terms of special functions
\beq
\xi = \frac{P_0}{2\pi }\int_{k_m}^\infty dk \, \frac{J_0(kr)}{k} = \frac{P_0}{4\pi}G^{23}_{01}\lp (k_m r/2)^2\rp,
\eeq
where $G$ is the Meijer G function. Defining a reduced area $a=k_{m}^2 r^2/2$ and consider separations that are small compared to the system size (corresponding to the scale of continents) $a \ll 1$, we can expand
\beq \xi(r) \approx \frac{P_0}{4\pi} \lp -\gamma +\frac{1}{2}\lp -\ln a +a -\frac{a^2}{8}+\frac{a^3}{108}\rp\rp
\eeq
where $\gamma \approx 5.7721$ is Euler's constant and we only neglect terms $\mathcal{O}{\lp(k_m r)^8\rp}$. The second term guarantees that $\xi \gg 1$ for sufficiently small $a$ and $\xi < 0$ for $a\gtrsim 0.51$. Most importantly, we note that for $r$ much smaller than the system size,
\beq
\label{log divergence}
\xi(r) \ra - \frac{P_0}{4\pi} \ln r.
\eeq
Since it is possible to invert a Fourier transform, any 2D model which predicts a correlation function that logarithmically diverges for small $r$ must have a power spectrum of the form $P \propto k^{-2}$ for $k \gg k_m$.
%%%%%%%%%%%%%%%%%%%%%%%%%%%%%%%%%%%%%%%%%%%%%%%%%%%%%%%%%%%%%%%%%%%%%%%%
\section{Effective field theory}
%%%%%%%%%%%%%%%%%%%%%%%%%%%%%%%%%%%%%%%%%%%%%%%%%%%%%%%%%%%%%%%%%%%%%%%%
The basic program of effective field theory is the following: given a statistical physics system in a fixed number of dimensions (in this case $D=2$), write down the Hamiltonian 
\beq H = \int d^2 x \, \mathcal{H}(\delta,\nabla \delta, \nabla\nabla \delta, \dots ), \eeq
such that $\mathcal{H}$ contains all terms which are consistent with the symmetries of the system. By universality, the macroscopic properties of the system should then be reflected in the field theory. For a pedagogical introduction to this approach, see \citep{kardar}. In our case, the symmetries are particularly constraining: we want the Hamiltonian to be invariant under scaling operations $x \ra \lambda x$ in addition to translations and rotations of the Euclidean plane. Under a change of scale, the population density transforms like a scalar, so $\delta(x) \ra \delta (\lambda^{-1} x)$ while $d^2 x \ra \lambda^2 d^2 x$ and $\nabla \delta \ra \lambda^{-1} \nabla \delta$. Hence scale invariance requires that each term in $\mathcal{H}$ contain exactly two derivatives to cancel the $\lambda^2$ from the area element. Rotational symmetry then limits us to only one possible term $(\nabla \delta)^2$:
\beq
H =  \frac{1}{2} \int d^2 x \,(\nabla \delta)^2,
\eeq
which is simply a free scalar field in two dimensions. Adding any interaction term to $\mathcal{H}$ of the form $V(\delta)$ is not allowed, as $d^2 x\, V(\delta)$ would not transform correctly under a scale. Using standard field theory techniques, one can show that the correlation function $\xi(r)$ has the form of (\ref{log divergence}); hence for wave vectors $k \gg k_m$ (corresponding to physical scales shorter than the system size), we must have that the power spectrum $P \propto k^{-2}$. This concludes our proof that a scalar random field in 2 spatial dimensions will have a power spectrum $P \propto k^{-2}$.

Let us make some further comments about (D2) that may be helpful to readers unfamiliar with effective field theory techniques. In particular, we can use (D2) to construct alternate theories that also will yield power spectra $P \propto k^{-2}$. For example, the generic Langevin equation, specialized to the case where the Hamiltonian is given by (D2), is just the famous diffusion equation with a noise term\citep{kardar}:
\beq
\frac{\partial \delta}{\partial t} = - \lambda \nabla^2 \delta + \eta,
\eeq
where $\delta$ is the fractional population over-density and $\eta$ is a fluctuating random variable. %This equation is the Langevin equation corresponding to the Hamiltonian
Notice that while this model also involves a diffusion equation, it is conceptually distinct from equation (2). Here we think of population as physically diffusing in Cartesian space; in equation (2), the diffusion is not happening in Cartesian space but in Fourier space. Since (D3) was derived from a Langevin equation corresponding to the Hamiltonian (D2), its equilibrium properties must be described by (D2); hence it follows that $P(k)\propto k^{-2}$. Note, however, that the noisy diffusion model lacks parameters dimensions of length to any power. Thus, our simple dimensional analysis argument presented after equation (3) still holds, and provides a simpler derivation of the power spectrum.
\end{appendix}
%\begin{small}
%\addcontentsline{toc}{section}{References}
%\bibliographystyle{apsrev4-1}
%\bibliography{biblio}
%\end{small}
\bibliography{biblio}{}

\end{document}